\documentclass[12pt,aps,nofootinbib]{revtex4-2}

\usepackage{ulem}
\usepackage{amsmath,amssymb,bm}
\usepackage[dvips]{graphicx}
\usepackage{hyperref}
\usepackage{yfonts}
\usepackage{color}

\newcommand{\beq}{\begin{eqnarray}}
\newcommand{\eeq}{\end{eqnarray}}
\renewcommand\d{\partial}

\begin{document}

\title{Helical magnetic effect and the chiral anomaly}

\author{Naoki Yamamoto$^1$ and Di-Lun Yang$^{1,2}$}
\affiliation{$^1$Department of Physics,  Keio University, Yokohama 223-8522, Japan \\
	$^2$Institute of Physics, Academia Sinica, Taipei, 11529, Taiwan}
\begin{abstract}
In the presence of the fluid helicity ${\bm v} \cdot {\bm \omega}$, the 
magnetic field induces an electric current of the form 
${\bm j} = C_{\rm HME} ({\bm v} \cdot {\bm \omega}) {\bm B}$.
This is the helical magnetic effect (HME). We show that for massless Dirac fermions 
with charge $e=1$, the transport coefficient $C_{\rm HME}$ is fixed by the 
chiral anomaly coefficient $C=1/(2\pi^2)$ as $C_{\rm HME} = C/2$
independently of interactions. We show the conjecture that the coefficient of the 
magnetovorticity coupling for the local vector charge, 
$n = C_{B \omega} {\bm B} \cdot {\bm \omega}$, 
is related to the chiral anomaly coefficient as $C_{B \omega} = C/2$. 
We also discuss the condition for the emergence of the helical plasma instability 
that originates from the HME.
\end{abstract}
\maketitle

\section{Introduction}
Chiral transport phenomena in relativistic chiral matter have attracted growing 
interests in various physical systems, such as heavy ion collisions \cite{Kharzeev:2015znc},
early Universe \cite{Joyce:1997uy}, core-collapse supernovae \cite{Yamamoto:2015gzz},
and Weyl semimetals \cite{Hosur:2013kxa}.
One prototype example is the chiral magnetic effect (CME) 
\cite{Vilenkin:1980fu,Nielsen:1983rb,Alekseev:1998ds,Fukushima:2008xe}---the 
electric current along a magnetic field in the presence of a chirality imbalance of 
fermions, characterized by a chiral chemical potential $\mu_5$.

In Ref.~\cite{Yamamoto:2015gzz}, it was pointed out that in the presence of a 
finite helicity, such as fluid helicity ${\bm v} \cdot {\bm \omega}$,%
\footnote{The fluid helicity has been known to have an important role in the hydrodynamics 
\cite{Moffatt1969}, and in particular, in the turbulence \cite{Yoshizawa}.} 
an electric current is induced by the magnetic field ${\bm B}$ even without $\mu_5$:
\beq
{\bm j} = C_{\rm HME} ({\bm v} \cdot {\bm \omega}) {\bm B}\,,
\eeq
where ${\bm v}$ is the local fluid velocity and ${\bm \omega} = \frac{1}{2} {\bm\nabla} \times {\bm v}$ 
is the vorticity. This was coined the helical magnetic effect (HME).%
\footnote{For the HME and helical vortical effects in other contexts, 
see Refs.~\cite{Kharzeev:2018jip,Ambrus:2019khr}.}
However, the transport coefficient $C_{\rm HME}$ has not been determined so far, 
largely because this is a nonlinear nonequilibrium transport in terms of external fields 
${\bm v}$, ${\bm \omega}$, and ${\bm B}$.

In this paper, we show that the transport coefficient $C_{\rm HME}$ for massless 
Dirac fermions is fixed by the coefficient of the chiral anomaly $C$ as
\beq
\label{C_HME}
C_{\rm HME} = \frac{C}{2}\,, \qquad C \equiv \frac{1}{2\pi^2}\,.
\eeq
Along the way, we also show the conjecture in Ref.~\cite{Hattori:2016njk} that the 
coefficient of the magnetovorticity coupling for the local vector charge,
\beq
n = C_{B \omega} {\bm B} \cdot {\bm \omega},
\eeq
is connected to the chiral anomaly coefficient $C$ as
\beq
\label{conjecture}
C_{B \omega} = \frac{C}{2}\,.
\eeq
While $C_{B \omega}=1/(4\pi^2)$ was derived for noninteracting massless Dirac 
fermions in the homogeneous magnetic field in Ref.~\cite{Hattori:2016njk}, our 
derivation shows that Eq.~(\ref{conjecture}) is exact even in the presence of 
interactions for generic inhomogeneous magnetic fields.
 
Our derivation is based on the idea that, under the assumption that the system of 
interest is in local thermal equilibrium, the vorticity is introduced as a fictitious 
spacetime torsion \cite{Shitade:2014iya,Khaidukov:2018oat}, which, in turn can be 
regarded as a background axial gauge field \cite{Shapiro:2001rz}. 
This is in spirit similar to Luttinger's argument that a temperature gradient can 
be introduced as a fictitious gravitational field \cite{Luttinger:1964zz}. In this way, 
we will show that the HME may be understood as a kind of the chiral torsional effect 
recently discussed in literature 
\cite{Khaidukov:2018oat,Imaki:2020csc,Ferreiros:2020uda,Manes:2020zdd}.

In Ref.~\cite{Yamamoto:2015gzz}, it was also argued that the presence of the HME 
leads to a new type of plasma instability, coined the helical plasma instability (HPI), 
in the same way that the CME induces the chiral plasma instability (CPI) \cite{Akamatsu:2013pjd}.
In this paper, we also discuss under which conditions the HPI can appear.

This paper is organized as follows. In Sec.~\ref{sec:CS}, we review the Chern-Simons 
currents for massless Dirac fermions. In Sec.~\ref{sec:vorticity}, we argue that the vorticity 
can be realized as an axial gauge field. In Sec.~\ref{sec:HME}, we show that the coefficients 
of the HMI and other related vorticity-induced effects are fixed by the anomaly coefficient. 
In Sec.~\ref{sec:HPI}, we discuss the condition for the emergence of the HPI. 
In Sec.~\ref{sec:discussion}, we make concluding remarks and discuss open questions.

Throughout the paper, we use the natural units $\hbar = c = k_{\rm B} = 1$. We absorb 
the elementary charge $e$ into the definition of the gauge field $A_{\mu}$ 
unless stated otherwise. We use the Minkowski metric $\eta_{\mu \nu} = (+1, -1, -1, -1)$ 
and define the totally antisymmetric tensor $\epsilon^{\mu \nu \alpha \beta}$ 
such that $\epsilon^{0123}=+1$. 
We also introduce the notations $A_{[\mu} B_{\nu]} \equiv A_{\mu} B_{\nu} - A_{\nu} B_{\mu}$ 
and $A_{(\mu} B_{\nu)} \equiv A_{\mu} B_{\nu} + A_{\nu} B_{\mu}$.

\section{Chern-Simons currents}
\label{sec:CS}
We first review the Chern-Simons currents for massless Dirac fermions that will be used 
in the following discussions; see, e.g., Ref.~\cite{Landsteiner:2016led} for a recent review. 
We consider a system of massless Dirac fermions coupled to the vector and axial gauge 
fields, $A^{\mu}$ and $A^{\mu}_5$:
\beq
{\cal L} = \bar \psi ({\rm i} \gamma^{\mu} \d_{\mu} - \gamma^{\mu} A_{\mu} - \gamma^{\mu} \gamma^5 A_{\mu}^5) \psi \,.
\eeq
One may alternatively regard this theory as right- and left-handed fermions coupled to 
right- and left-handed gauge fields, $A^{\mu}_{\rm R} \equiv A^{\mu} + A^{\mu}_5$ 
and $A^{\mu}_{\rm L} \equiv A^{\mu} - A^{\mu}_5$, respectively, 
\beq
{\cal L} = \psi^{\dag}_{\rm R} {\rm i} \sigma^{\mu} (\d_{\mu} + {\rm i} A_{\mu}^{\rm R}) \psi_{\rm R}
+ \psi^{\dag}_{\rm L} {\rm i} \bar \sigma^{\mu} (\d_{\mu} + {\rm i} A_{\mu}^{\rm L}) \psi_{\rm L}\,,
\eeq
where $\psi_{\rm R,L}$ are right- and left-handed fermions, $\sigma^{\mu} = (1, {\bm \sigma})$, 
and $\bar \sigma^{\mu} = (1, -{\bm \sigma})$ with $\sigma^i$ ($i=1,2,3$) being the Pauli matrices.
One can then derive the covariant anomalies for the right- and left-handed sector as
\beq
\d_{\mu} j^{\mu}_{\chi} = \mp \frac{C}{16} \epsilon^{\mu \nu \alpha \beta} F_{\mu \nu}^{\chi} F_{\alpha \beta}^{\chi} \qquad (\chi = {\rm R, L})\,,
\eeq
where $F_{\mu \nu}^{\chi} = \d_{\mu} A_{\nu}^{\chi} - \d_{\nu} A_{\mu}^{\chi}$.
By adding and subtracting the right- and left-handed sectors, we have
\begin{align}
\label{dj_0}
\d_{\mu} j^{\mu} &= -\frac{C}{4} \epsilon^{\mu \nu \alpha \beta} F_{\mu \nu} F_{\alpha \beta}^5\,, \\
\label{dj5_0}
\d_{\mu} j^{\mu}_5 &= -\frac{C}{8} \epsilon^{\mu \nu \alpha \beta} \left(F_{\mu \nu} F_{\alpha \beta} + F_{\mu \nu}^5 F_{\alpha \beta}^5 \right)\,,
\end{align}
where $F_{\mu \nu} = \d_{\mu} A_{\nu} - \d_{\nu} A_{\mu}$ and 
$F_{\mu \nu}^{5} = \d_{\mu} A_{\nu}^{5} - \d_{\nu} A_{\mu}^{5}$.

We will be interested in the case where $A_{\mu}$ is dynamical while $A_{\mu}^5$ is 
external. In this case, Eq.~(\ref{dj_0}) is problematic in the sense that it is inconsistent 
with the gauge symmetry. In fact, the equation of motion for $A_{\mu}$ is given by 
Maxwell's equations
\beq
\d_{\nu} F^{\nu \mu} = j^{\mu},
 \eeq
and so
\beq
\label{dj_1}
\d_{\mu} j^{\mu} = \d_{\mu} \d_{\nu} F^{\nu \mu} = 0.
\eeq
One then finds that Eq.~(\ref{dj_0}) is inconsistent with Eq.~(\ref{dj_1}).

The way out is well known. We can add a topological current called the Chern-Simons current
\beq
\label{CS}
j_{\rm CS}^{\mu} = \frac{C}{2}  \epsilon^{\mu \nu \alpha \beta} A_{\nu}^5 F_{\alpha \beta}
\eeq
into $j^{\mu}$ such that the right-hand side of Eq.~(\ref{dj_0}) is cancelled out 
(namely, $\d_{\mu} \tilde j^{\mu} = 0$, where $\tilde j^{\mu} = j^{\mu} + j^{\mu}_{\rm CS}$).
The temporal and spatial components read
\begin{align}
\label{n_CS}
n_{\rm CS} &= C {\bm A}^5 \cdot {\bm B}\,,
\\
\label{j_CS}
{\bm j}_{\rm CS} &= C A_0^5 {\bm B} + C {\bm E} \times {\bm A}_5\,,
\end{align}
respectively. Note that all the coefficients appearing in Eqs.~(\ref{n_CS}) and (\ref{j_CS})
are fixed by the anomaly coefficient $C$ by construction.

One might wonder what the physical realizations of the axial gauge field $A_{\mu}^5$ are.
There are in fact systems where $A_{\mu}^5$ appears emergently.
For example, in Weyl semimetals, 
$A_0^5 = b_0$ and ${\bm A}^5 = {\bm b}$, where $b_0$ and ${\bm b}$ 
correspond to the energy and momentum separations between two Weyl nodes.
As a result, the electric current at finite $\mu_5$ is given by
\beq
{\bm j} = C (\mu_5 + b_0) {\bm B} + C {\bm E} \times {\bm b} \,,
\eeq
where the first term is the CME \cite{Basar:2013iaa,Landsteiner:2013sja} and 
the second is the anomalous Hall effect \cite{Yang:2011,Grushin:2012mt,Zyuzin:2012tv,Goswami:2012db}.%
\footnote{In the context of Weyl semimetals, the importance of the Chern-Simons
contributions is stressed in Refs.~\cite{Gorbar:2016ygi,Gorbar:2017vph}.}
In particular, in equilibrium where $\mu_5 = -b_0$, the CME vanishes
as is consistent with the generalized Bloch theorem \cite{Yamamoto:2015fxa}.

We will next argue that the vorticity can also be understood as an emergent 
axial gauge field.

\section{Vorticity as an axial gauge field}
\label{sec:vorticity}
We are interested in the hydrodynamic regime of a gauge theory with finite vorticity 
${\bm \omega} \neq {\bm 0}$. Below we will use the following two facts: 
(i) the vorticity is introduced as a fictitious spacetime torsion \cite{Shitade:2014iya,Khaidukov:2018oat}, 
and (ii) the torsion can be regarded as an axial gauge field \cite{Shapiro:2001rz}.
As a result, the vorticity can be realized as an axial gauge field.

To see the fact (ii) first, consider a spacetime with torsion defined by
$T^{\rho}_{\mu \nu} \equiv \tilde \Gamma^{\rho}_{[\mu \nu]}$, where
$\tilde \Gamma^{\rho}_{\mu \nu}$ is a nonsymmetric affine connection satisfying 
$\tilde \Gamma^{\rho}_{\mu \nu} = \Gamma^{\rho}_{\mu \nu} + K^{\rho}_{\mu \nu}$,
with $\Gamma^{\rho}_{\mu \nu}$ being the symmetric Christoffel symbol and 
$K_{\mu \nu \rho} \equiv \frac{1}{2} (T_{\mu \nu \rho} - T_{\nu \mu \rho} - T_{\rho \mu \nu})$
the contorsion tensor.
The action for massless Dirac fermions is \cite{Shapiro:2001rz}
\beq
S = \int {\rm d}^4 x \sqrt{-g} \frac{\rm i}{2} \left(\bar \psi \gamma^{\mu} \tilde \nabla_{\mu} \psi - \tilde \nabla_{\mu} \bar \psi \gamma^{\mu} \psi \right)\,,
\eeq
where $g$ is the determinant of the spacetime metric and
\begin{gather}
\tilde \nabla_{\mu} \psi = \d_{\mu} \psi + \frac{\rm i}{4} \tilde \omega_{\mu}^{\hat a \hat b} \sigma_{\hat a \hat b} \psi\,, \qquad
\tilde \nabla_{\mu} \bar \psi = \d_{\mu} \bar \psi - \frac{\rm i}{4} \bar \psi \tilde \omega_{\mu}^{\hat a \hat b} \sigma_{\hat a \hat b}\,,
\\
\tilde \omega_{\mu \hat a \hat b} = \omega_{\mu \hat a \hat b} + K^{\alpha}_{\cdot \lambda \mu} e^{\lambda}_{\hat a} e_{\hat b \alpha}\,.
\end{gather}
Here, 
$\omega_{\mu \hat a \hat b} = e_{\hat b \alpha} \d_{\mu} e^{\alpha}_{\hat a} + \Gamma^{\alpha}_{\lambda \mu} e^{\lambda}_{\hat a} e_{\hat b \alpha}$
is the spinor connection in the spacetime without torsion, $\sigma_{\hat a \hat b} = \frac{\rm i}{2}[\gamma_{\hat a}, \gamma_{\hat b}]$
with $\gamma^{\hat a}$ being the usual $\gamma$ matrix in flat spacetime, and 
$\gamma^{\mu} = e^{\mu}_{\hat a} \gamma^{\hat a}$ with $e^{\mu}_{\hat a}$ 
the vierbein satisfying $e^{\hat a}_{\mu} e^{\hat b \mu} = \eta^{\hat a \hat b}$ 
and $e^{\hat a}_{\mu} e^{\hat a}_{\nu} = g_{\mu \nu}$.
In the following, we will be interested in the torsional effects in flat spacetime. 
In this case, the Lagrangian can be rewritten as \cite{Shapiro:2001rz}
\beq
{\cal L} = \bar \psi \left({\rm i}\gamma^{\mu} \d_{\mu} - \frac{1}{8} \gamma^{\mu}\gamma^5 S_{\mu} \right) \psi\,,
\eeq
where $S_{\nu} \equiv \epsilon_{\alpha \beta \mu \nu} T^{\alpha \beta \mu}$.

To see the fact (i) above, we consider the metric
\beq
\label{g}
{\rm d}s^2 = {\rm d} t^2 + 2 {\bm v} \cdot {\rm d} {\bm x} {\rm d} t  - {\rm d} {\bm x}^2
\eeq
up to $O({\bm v}^1)$.
We here take ${\bm v}$ to depend on the coordinate ${\bm x}$ (but not on time $t$), 
${\bm v} = {\bm v}({\bm x})$, and in the following, we will focus on the terms related to 
the vorticity ${\bm \omega} = \frac{1}{2} {\bm\nabla} \times {\bm v} \neq {\bm 0}$ in
the local rest frame of the fluid.%
\footnote{The same setup is considered to derive the transport coefficient 
of the chiral vortical effect (CVE) in Ref.~\cite{Landsteiner:2011cp}.}
The vierbein corresponding to the metric (\ref{g}) is given by \cite{Khaidukov:2018oat}
\beq
\label{e}
e^{\hat 0}_0 = 1, \quad e^{\hat 0}_i = -v_i, \quad 
e^{\hat i}_0 = 0, \quad 
e^{\hat i}_j = \delta^{\hat i}_j.
\eeq
We now impose vierbein postulate 
$\tilde \nabla_{\mu} e^{\hat a}_{\nu} = 0$ 
for $\omega_{\mu \hat a \hat b} = 0$.  
From the antisymmetric and symmetric parts of this postulate with respect to the indices 
$\mu$ and $\nu$, we obtain
\begin{gather}
T^{\hat a}_{\mu \nu} = \d_{[\mu} e^{\hat a}_{\nu]}\,, 
\\
\label{constraint}
\d_{(\mu} e_{\nu)}^{\hat a} = 2 \Gamma_{\mu \nu}^{\lambda} e^{\hat a}_{\lambda} - T_{(\mu \lambda \nu)}e^{{\hat a} \lambda}\,,
\end{gather}
where $T^{\hat a}_{\mu \nu} = T^{\rho}_{\mu \nu} e^{\hat a}_{\rho}$.
For the specific choice of the vierbein in Eq.~(\ref{e}), we have
\begin{align}
T^{\hat 0}_{i j} &= - \d_{[i} v_{j]} =  -2 \epsilon_{ijk} \omega^k \,,
\label{T}
\end{align}
where $\epsilon_{ijk} \equiv \epsilon_{0ijk}$ and the other components of 
$T^{\hat a}_{\mu \nu}$ vanish.
One can also check that the $(\hat a, \mu, \nu) = (\hat 0, 0, i), (\hat 0, i, 0)$ 
components of Eq.~(\ref{constraint}) lead to the constraint $\partial_t {\bm v} = {\bm 0}$ 
while Eq.~(\ref{constraint}) is automatically satisfied for the other components 
at the order of $O({\bm v}^1)$. Consequently, we have
\beq
S_{\mu} = (0, 4 \omega_i).
\eeq

So far, we have considered the local rest frame of the fluid. 
The expression of $S^{\mu}$ in the generic inertial frame, where the local fluid four velocity 
is $u^{\mu} = \gamma(1, {\bm v})$ with $\gamma=(1-{\bm v}^2)^{-1/2}$,
can be obtained by performing a Lorentz boost as
\beq
S_{\mu} = 4 \omega_{\mu},
\eeq
where $\omega_{\mu} = \frac{1}{2} \epsilon_{\mu \nu \alpha \beta} u^{\nu} \d^{\alpha} u^{\beta}$.

In summary, we see that the vorticity can be introduced as an {\it emergent} background 
axial gauge field
\beq
\label{A5}
A_{\mu}^5 \equiv \frac{1}{8} S_{\mu}  = \frac{1}{2} \omega_{\mu}
\eeq
in the generic inertial frame.

\section{Helical magnetic effect and other vorticity-induced effects}
\label{sec:HME}
We are now ready to derive the vorticity-induced effects in the hydrodynamic regime 
of a gauge theory by combining the results in the previous sections.

By inserting Eq.~(\ref{A5}) into Eq.~(\ref{CS}), we find
\beq
\label{main}
j^{\mu}_{\rm CS} = \frac{C}{4} \epsilon^{\mu \nu \alpha \beta} \omega_{\nu} F_{\alpha \beta}\,.
\eeq
We can also rewrite Eq.~(\ref{main}) in terms of the electromagnetic fields defined 
in the fluid rest frame,
$E^{\mu} = F^{\mu \nu} u_{\nu}$ and 
$B^{\mu} = \frac{1}{2} \epsilon^{\mu \nu \alpha \beta} u_{\nu} F^{\alpha \beta}$.
By using the decomposition
\beq
F_{\mu \nu} = E_{(\mu}u_{\nu)} + \epsilon_{\mu \nu \alpha \beta} u^{\alpha} B^{\beta}
\eeq
and the identity $u \cdot \omega = 0$, we obtain
\beq
\label{j_cov}
j^{\mu}_{\rm CS} = -\frac{C}{2} (B \cdot \omega) u^{\mu} - \frac{C}{2} \epsilon^{\mu \nu \alpha \beta} u_{\nu} E_{\alpha} \omega_{\beta}\,.
\eeq
In the local rest frame where $u^{\mu}=(1,{\bm 0})$, we have
\begin{gather}
\label{MV}
n_{\rm CS} = \frac{C}{2} {\bm B} \cdot {\bm \omega}\,,
\\
\label{Eomega}
{\bm j}_{\rm CS} = \frac{C}{2} {\bm E} \times {\bm \omega}\,.
\end{gather}
Equations~(\ref{j_cov}) and (\ref{MV}) agree with the results obtained by different approaches
based on the Wigner function formalism \cite{Yang:2020mtz} and the Landau level picture 
\cite{Hattori:2016njk,Lin:2021sjw}, respectively. 
We however emphasize that while the results of Refs.~\cite{Hattori:2016njk,Yang:2020mtz,Lin:2021sjw} 
were obtained for free Dirac fermions in the homogenous electromagnetic field, 
our derivation extends it to the case of interacting Dirac fermions in the generic 
inhomogeneous electromagnetic field. In particular, this derivation shows that 
the transport coefficients of Eqs.~(\ref{main}), (\ref{j_cov}), (\ref{MV}), and (\ref{Eomega}) 
are completely fixed by the anomaly coefficient $C$ independently of interactions. 
This proves the conjecture of Ref.~\cite{Hattori:2016njk} that the coefficient of 
the magnetovorticity coupling in Eq.~(\ref{MV}) is related to $C$.
Equation~(\ref{Eomega}) also indicates a new type of anomaly-related current, which does 
not require the presence of $\mu_5$ unlike the CME or other chiral transport phenomena.
This is a dissipationless current which does not generate any entropy because 
${\bm j} \cdot {\bm E} = 0$.

Let us now consider the case with $|{\bm v}| \ll 1$ in the inertial frame, for which $u^{\mu} \approx (1, {\bm v})$.
In this case, we have the following contribution to the electric current in addition to Eq.~(\ref{Eomega}):%
\footnote{If $\d_t {\bm v} \neq {\bm 0}$, we would also have an analog of the anomalous Hall effect 
$\Delta {\bm j}_{\rm CS} = \frac{C}{4} {\bm E} \times ({\bm v} \times \d_t {\bm v})$. However, in the 
present derivation, this term is absent because of the constraint $\d_t {\bm v} = {\bm 0}$.}
\begin{align}
\Delta {\bm j}_{\rm CS} = \frac{C}{2}({\bm v} \cdot {\bm \omega}) {\bm B} 
+ O({\bm v}^3) \,.
\label{HME}
\end{align}
This is the HME. Again, our derivation shows that the transport coefficient of the HME 
is fixed by the anomaly coefficient $C$ and is exact independently of interactions.

We note that the expression of the HME is similar to that of the CME
\cite{Vilenkin:1980fu,Nielsen:1983rb,Alekseev:1998ds,Fukushima:2008xe},
${\bm j} = C {\mu_5}{\bm B}$,
where $\mu_5 \equiv (\mu_{\rm R} - \mu_{\rm L})/2$ is the chiral chemical potential
with $\mu_{\rm R, L}$ being the chemical potentials for right- and left-handed fermions.
We can see that the correspondence between the HME and CME is 
${\bm v} \cdot {\bm \omega} \leftrightarrow \mu_{\rm R} - \mu_{\rm L}$.
While it is natural to have such a correspondence since ${\bm v} \cdot {\bm \omega}$ 
has the same quantum number as $\mu_{\rm R} - \mu_{\rm L}$ \cite{Yamamoto:2015gzz},
whether the prefactors are exactly the same is {\it a priori} nontrivial. 
We here show that this is indeed the case by revealing the relation of the HME 
to the chiral anomaly. 
Note also that the HME can be present even when $\mu_5 = 0$ unlike the CME.

\section{Helical plasma instability}
\label{sec:HPI}
One of the consequences of the HME is a new type of plasma instability, called 
the helical plasma instability \cite{Yamamoto:2015gzz}.
Here, let us discuss the condition for the emergence of the HPI in details.

We consider a system that has a finite fluid helicity $n_{\rm flu}(\bm x) = {\bm v} \cdot {\bm \omega}$
in some region. For simplicity, we assume that $|{\bm v}| \ll 1$ and the spatial variation of 
the fluid helicity is sufficiently small. In order to focus on the consequences of the HME, 
we set $\mu_5 = 0$.

\subsection{Dissipationless fluids}
\label{sec:dissipationless}
We first consider the ideal situation of relativistic fluids without any dissipation.
Gauss's law and Amp\`ere's law including the Chern-Simons current read
\begin{gather}
{\bm \nabla} \cdot {\bm E} = {n}_{\rm CS} + n_{\rm back},
\\
{\bm \nabla} \times {\bm B} = {\bm j}_{\rm CS} + \d_t {\bm E},
\end{gather}
where $n_{\rm back}$ is the background charge density and 
${\bm j}_{\rm CS}$ is given by Eqs.~(\ref{Eomega}) and (\ref{HME}).
Below we assume the local charge neutrality ${n}_{\rm CS} + n_{\rm back} = 0$.
Combining these equations with Faraday's law 
${\bm \nabla} \times {\bm E} = - \partial_t {\bm B}$ and 
${\bm \nabla} \cdot {\bm B} = 0$, we obtain
\beq
\label{B}
(\d_t^2 - {\bm \nabla}^2) {\bm B} 
= \sigma_{\rm eff} {\bm \nabla} \times {\bm B}
+ \frac{C}{2} ({\bm \omega} \cdot {\bm \nabla}) {\bm E}\,,
\eeq
where we defined
\beq
\sigma_{\rm eff} \equiv C_{\rm HME} n_{\rm flu} = \frac{n_{\rm flu}}{4\pi^2}\,
\eeq
and assumed that spatial variation of ${\bm \omega}$ is sufficiently small.

To see that Eq.~(\ref{B}) has an unstable mode, consider a specific configuration,
$v_x \neq 0$, $v_y=v_z\approx 0$ and $\d_y v^z=-\d_z v^y=\omega^x\neq 0$, 
such that $n_{\rm flu} = v^x \omega^x \neq 0$, in the region of interest. 
We then seek for a solution of the gauge field ${\bm A}$ in the helicity basis as
\beq
{\bm A}_{\pm} = ({\bm e}_x \pm {\rm i} {\bm e}_y) {\rm e}^{-{\rm i} \omega t + {\rm i} k z},
\eeq
where ${\bm e}_{x,y}$ are the unit vectors in the $x,y$ directions and the subscript $\pm$ 
denotes the helicity $h=\pm 1$ states for $k > 0$. Here, we chose the temporal gauge 
$A_t = 0$ without loss of generality. It then follows that
\beq
{\bm E}_{\pm} = {\rm i} \omega {\bm A}_{\pm}, \qquad 
{\bm B}_{\pm} = \pm k {\bm A}_{\pm}.
\eeq
In this setup, the second term on the right-hand side of Eq.~(\ref{B}) vanishes, and 
the dispersion relations for ${\bm B}_{\pm}$ become
\beq
\omega^2 = k(k \mp \sigma_{\rm eff}).
\eeq
When $n_{\rm flu} > 0$ and $n_{\rm flu} < 0$, $\omega$ has the positive 
imaginary part in the region $0 < k < |\sigma_{\rm eff}|$ for the $h = 1$ and $h = -1$ states, 
respectively. 
In particular, this imaginary part becomes maximal at $k_{\rm inst} = |\sigma_{\rm eff}|/2$,
for which the time evolution of the magnetic field is given by the exponentially growing 
behavior with the maximum exponent,
\beq
B(t) = B(0) {\rm e}^{|\sigma_{\rm eff}| t/2}.
\eeq
Here $B(t)$ is the magnitude of the magnetic field at time $t$.
Therefore, the typical length and timescales of the HPI in this case are
\beq
\label{instability}
{\ell}_{\rm inst} = t_{\rm inst} = \frac{8 \pi^2}{|n_{\rm flu}|}\,.
\eeq

Similarly to the CPI, the magnetic field generated as a consequence of this HPI 
has finite helicity (positive helicity for $n_{\rm flu} > 0$ and negative helicity for 
$n_{\rm flu} < 0$), which is characterized by the magnetic helicity
\beq
h_{\rm mag} \equiv \int {\rm d}^3 {\bm x} \ {\bm A} \cdot {\bm B}\,.
\eeq
In other words, the HPI is a dynamical process that converts the fluid helicity to 
the magnetic helicity and the fluid kinetic energy to the magnetic energy such that
the total helicity and total energy are conserved \cite{Yamamoto:2015gzz}.

\subsection{Dissipative fluids}
\label{sec:dissipative}
We then consider the case of dissipative fluids and take into account the contribution 
of the Ohmic current. In this case, the electric current is given by
\beq
{\bm j} = \sigma ({\bm E} + {\bm v} \times {\bm B} ) 
+ \frac{C}{2}\left[({\bm v} \cdot {\bm \omega}){\bm B} + {\bm E} \times {\bm \omega} \right]\,,
\eeq
where $\sigma$ is the electrical conductivity.
Combining it with Amp\`ere's law ${\bm \nabla} \times {\bm B} = {\bm j}$
(where the displacement current $\d_t {\bm E}$ can be ignored; see footnote~6), 
we have
\beq
{\bm E} = - {\bm v} \times {\bm B} + \eta {\bm \nabla} \times {\bm B} 
- \frac{C \eta}{2} \left[({\bm v} \cdot {\bm \omega}){\bm B} + {\bm E} \times {\bm \omega} \right]\,,
\eeq
where $\eta \equiv 1/\sigma$ is the resistivity. 
When $\eta |{\bm \omega}| \ll 1$, this equation can be solved in terms of ${\bm E}$ as
\beq
\label{E}
{\bm E} = - {\bm v} \times {\bm B} + \eta {\bm \nabla} \times {\bm B} 
- \frac{C \eta}{2} ({\bm \omega} \cdot {\bm B}){\bm v} 
+ \frac{C \eta^2}{2}\left[{\bm \nabla} ({\bm \omega} \cdot {\bm B})- ({\bm \omega} \cdot {\bm \nabla}) {\bm B} \right]\,,
\eeq
where we used 
$({\bm v} \times {\bm B}) \times {\bm \omega} - ({\bm v} \cdot {\bm \omega}){\bm B} = - ({\bm \omega} \cdot {\bm B}){\bm v}$.
Note here that the term involving the fluid helicity disappears due to the cancellation between the 
contributions from the currents $({\bm v} \cdot {\bm \omega}){\bm B}$ and ${\bm E} \times {\bm \omega}$
with the same transport coefficients.
The terms with the coefficient $C \eta^2/2$ in Eq.~(\ref{E}) are higher order in derivatives 
and will be ignored below. By inserting Eq.~(\ref{E}) into Faraday's law, we obtain
\beq
\d_t {\bm B} = {\bm \nabla} \times ({\bm v} \times {\bm B}) + \eta{\bm \nabla}^2 {\bm B} 
+ \frac{C \eta}{2} {\bm \nabla} \times \left[({\bm \omega} \cdot {\bm B}){\bm v} \right]\,.
\eeq
In this case, since there is no parity-violating term involving fluid helicity, 
it does not exhibit the HPI unlike Eq.~(\ref{B}).

\subsection{Condition for the helical plasma instability}
The origin of this qualitative difference of dissipative fluids from dissipationless ones 
in Sec.~\ref{sec:dissipative} is that the electric field is a dependent variable of the 
magnetic field as given by Eq.~(\ref{E}) and it can be integrated out. 
In terms of the notion of generalized global symmetries \cite{Gaiotto:2014kfa}, 
this stems from the fact that the electric one-form symmetry is explicitly broken in medium 
and the electric field is no longer a low-energy degree of freedom in the usual 
formulation of magnetohydrodynamics \cite{Grozdanov:2016tdf,Glorioso:2018kcp}.
What distinguishes the two scenarios is whether the  the typical timescale of the 
HPI, $t_{\rm inst}$ in Eq.~(\ref{instability}), is sufficiently small compared with 
the inverse of the gap of the electric field, given by $\eta$:%
\footnote{The fact that the electric field has a gap $\sigma$ can be seen as follows \cite{Davidson}:
by substituting the Ohmic law ${\bm j} = \sigma {\bm E}$ into the continuity equation
$\d_t n + {\bm \nabla} \cdot {\bm j} = 0$ and by using the Gauss law 
${\bm \nabla} \cdot {\bm E} = n$, one obtains $\d_t n + \sigma n = 0$.
Therefore, $n \propto {\rm e}^{- \sigma t}$, showing the presence of the gap $\sigma$.
We also have $|\d_t {\bm E}| = \eta |\d_t {\bm j}| \ll |{\bm j}|$ for the timescale 
$t \gg \eta$, which justifies the assumption that the displacement current in the Amp\`ere's 
law is negligible.}
the HPI becomes relevant when $t_{\rm inst} \ll \eta$, or equivalently, 
$n_{\rm flu} \gg \sigma$.

In the case of weakly coupled QED plasma at finite temperature $T$, 
the mean free path ${\ell}_{\rm mfp}$ and conductivity $\sigma$ are 
given parametrically by \cite{Baym:1990uj,Arnold:2000dr}
\beq
{\ell}_{\rm mfp} \sim \frac{1}{e^4 T}\,, \qquad
\sigma \sim e^2 T^2 {\ell}_{\rm mfp} \sim \frac{T}{e^2}\,,
\eeq
except for logarithmic correction, where we restored the coupling constant $e$.
On the other hand, $n_{\rm flu}$ is estimated as 
\beq
n_{\rm flu} \sim \frac{v^2}{L}\,,
\eeq
where $L$ is the typical length scale for the variation of the hydrodynamic 
variables and $v$ is the typical magnitude of the fluid velocity.
In the hydrodynamic regime where $L \gg \ell_{\rm mfp}$, 
we have $n_{\rm flu} \ll \sigma$ at the weak coupling $e \ll 1$, 
and hence, there is no HPI.

Note however that the discussion so far is limited to the weakly coupled plasma
with massless Dirac fermions in the hydrodynamic regime. The following are two
examples where this discussion does not apply and the HPI can appear:
\begin{itemize}
\item{Strongly coupled plasma: in the case of quark-gluon plasmas (QGP), 
the parametric dependence of $\sigma$ is changed to \cite{Arnold:2000dr}
\beq\label{sigma_QCD}
{\ell}_{\rm mfp} \sim \frac{1}{g^4 T}\,, \qquad
\sigma \sim e^2 T^2 {\ell}_{\rm mfp} \sim \frac{e^2T}{g^4}\,,
\eeq
where $g$ is the QCD coupling constant. Although this estimate is obtained by the 
weak-coupling analysis and is applicable to $g \ll 1$ strictly speaking, if we extrapolate
this formula to $g \gtrsim 1$, there is a regime where $n_{\rm flu} \gg \sigma$ is satisfied. 
To see this, we rewrite the mean free path as 
$\ell_{\rm mfp}\sim e^2/(g^8\sigma)$ from Eq.~(\ref{sigma_QCD}). 
The hydrodynamic condition $L \gg {\ell}_{\rm mfp}$ then yields 
$n_{\rm flu}\ll v^2g^8\sigma/e^2$. To satisfy the condition for HPI simultaneously, 
it is required that $g^4\gg e/v$, which could be satisfied in strongly coupled QCD, 
but can never be satisfied in QED, for which $g$ is replaced by $e$.
Therefore, the HPI can emerge in the strongly coupled QGP with a finite fluid helicity.}
\item{Nonrelativistic corrections: the fact that both of transport coefficients of the currents
$({\bm v} \cdot \bm \omega) {\bm B}$ and ${\bm E} \times {\bm \omega}$ are given by
the same factor related to the chiral anomaly, $C/2$, is specific to the case for massless 
Dirac fermions and is not generically applicable to the case for massive Dirac fermions
and nonrelativistic particles. Since these two currents are related to each other via 
Lorentz transformation as can be understood from Eq.~(\ref{j_cov}), the cancellation of 
the two should not be complete, e.g., for a nonrelativistic particle. 
In this case, the remaining term proportional to $({\bm v} \cdot {\bm \omega}) {\bm B}$ 
leads to the HPI in a way similar to Eq.~(\ref{B}).
Such a case is relevant to core-collapse supernovae (see below).}
\end{itemize}

\section{Discussions and outlook}
\label{sec:discussion}
In this paper, we have shown that the coefficients of the HME and the magnetovorticity 
coupling are completely fixed by the anomaly coefficient.
Although the HME looks similar to the CME, the prominent feature of the HME is that
it exists even when $\mu_5 = 0$. This is phenomenologically important because 
while $\mu_5$ is attenuated by a small but finite fermion mass, which could 
then suppress the CME, e.g., in core-collapse supernovae \cite{Grabowska:2014efa} and 
the electroweak plasma in the early Universe \cite{Boyarsky:2020cyk,Boyarsky:2020ani},
the fluid helicity is not. 

In the context of core-collapse supernovae, fluid helicity can be generated through 
the CVE of the neutrino in local thermal equilibrium, 
${\bm j} = -\left(\frac{\mu_{\nu}^2}{4\pi^2} + \frac{T^2}{12} \right){\bm \omega}$ 
with $\mu_{\nu}$ the neutrino chemical potential \cite{Vilenkin:1979ui,Landsteiner:2011cp},
as pointed out in Ref.~\cite{Yamamoto:2015gzz}.%
\footnote{We note that the generation of fluid helicity is not necessarily limited to the 
situation where neutrinos are in local thermal equilibrium. In fact, it has been recently 
shown using the chiral kinetic theory for neutrinos coupled to hydrodynamics for matter 
(dubbed the chiral radiation transfer theory) that the neutrino-matter collision can generate 
fluid helicity of matter even when neutrinos are away from equilibrium \cite{Yamamoto:2020zrs}.}
Then, the resulting fluid helicity induces the HME for relativistic electrons and 
nonrelativistic protons, and the former transport coefficient is given by Eq.~(\ref{C_HME}) 
(with a possible small mass correction). While the HME for electrons presumably does not 
lead to HPI, the HME for protons may lead to HPI as argued above.
This could be a potential mechanism for the inverse energy cascade in core-collapse 
supernovae \cite{Masada:2018swb}. It would also be interesting to study the possible 
helical instability beyond the hydrodynamic regime of neutrinos.

How efficiently the fluid helicity is generated in these systems is a nonlinear problem and 
should be investigated numerically by the helical magnetohydrodynamics incorporating 
the helical effects \cite{Yamamoto:2015gzz}. To this end, one first needs to extend the helical 
magnetohydrodynamics to the second order by including the HME and other possible effects.
It is especially important to explore the possible anomaly-related corrections to 
the energy-momentum tensor at the second order.

From the theoretical viewpoints, it would be interesting to study how the topological 
quantization of the transport coefficient of the HME and that of the magnetovorticity coupling 
can be understood in terms of the Berry curvature in a way similar to that of the CME 
\cite{Son:2012wh,Stephanov:2012ki,Son:2012zy}.
It would also be interesting to derive these second-order helical transport phenomena 
from the underlying quantum field theory based on the Wigner function formalism, such as 
Refs.~\cite{Hidaka:2016yjf,Hidaka:2017auj,Hayata:2020sqz}, beyond the homogeneous
electromagnetic fields considered in Ref.~\cite{Yang:2020mtz}.

Finally, we note that we assumed massless Dirac fermions in this paper. 
When Dirac fermions dynamically acquire a finite mass via interactions, 
the 't Hooft anomaly matching condition requires that there must be other gapless 
modes in the system that are responsible for the chiral anomaly. In this case, one expects 
that the HME is carried by these gapless modes just as the CME and CVE are carried by 
Nambu-Goldstone modes \cite{Fukushima:2012fg,Huang:2017pqe}. Accordingly, there 
must be an analog of the Wess-Zumino-Witten term \cite{Wess:1971yu,Witten:1983tw} 
for the HME. This question is deferred to future work.

\section*{Acknowledgement}
We thank Koichi~Hattori for discussions and Taro~Kimura for comments.
This work was supported by the Keio Institute of Pure and Applied Sciences (KiPAS) project 
at Keio University and JSPS KAKENHI Grants No.~19K03852 and No.~20K14470.

\end{document}